\documentclass[11pt,letter]{article}

\usepackage{poma_style}

\usepackage{graphicx}
\usepackage{amsmath,amsfonts,bm,amsthm}
\usepackage{courier}
\usepackage[colorlinks=true,urlcolor=blue,linkcolor=black,citecolor=black]{hyperref}
\usepackage{color,pstricks} 
\usepackage{caption}
\usepackage{tipa}
\usepackage{amssymb}
\usepackage{xcolor}
\usepackage{natbib} 
\usepackage{array}
\usepackage[mathlines,running]{lineno}
\usepackage{setspace}

\newcommand{\comment}[1]{} 

\newcommand{\SNRninty}{SNR$_{90}$}

\title{The role of cue enhancement and frequency fine-tuning in hearing impaired phone recognition}
\author{Ali Abavisani and Mark A Hasegawa-Johnson\\
\small{University of Illinois at Urbana-Champaign}\\
\small{Dept.~of Electrical and Computer Engineering}\\
\small{The Beckman Inst.~405~N~Mathews~Ave~Urbana IL~61801}
\date{}
}

\begin{document}

\maketitle


\begin{abstract}

A speech-based hearing test is designed to identify the susceptible error-prone phones for individual hearing impaired (HI) ear. Only robust 
tokens in the experiment noise levels had been chosen for the test. The noise-robustness of tokens is measured as \SNRninty~of the token, 
which is 
the signal to the speech-weighted noise ratio where a normal hearing (NH) listener would recognize the token with an accuracy of 90\% on 
average. 
Two sets of 
tokens T$_1$ and T$_2$ having the same consonant-vowels but different talkers with distinct \SNRninty~had been presented with flat gain at 
listeners' most 
comfortable level. We studied the effects of frequency fine-tuning of the primary cue by presenting tokens of the same consonant but different 
vowels with similar \SNRninty. Additionally, we investigated the role of changing the intensity of primary cue in HI phone recognition, by 
presenting tokens from both sets T$_1$ and T$_2$. On average, 92\% of tokens are improved when we replaced the CV with the same CV but with a 
more 
robust talker. Additionally, using CVs with similar \SNRninty~, on average, tokens are improved by 75\%, 71\%, 63\%, and 72\%, when we replaced 
vowels \textipa{/A, \ae, I, E/}, respectively. The confusion pattern in each case provides insight into how these changes affect the phone 
recognition in each HI 
ear. We propose to prescribe hearing aid amplification tailored to individual HI ears, based on the confusion pattern, the response from cue 
enhancement, and the response from frequency fine-tuning of the cue.

\end{abstract}

\section{Introduction}

World health organization statistics shows that over 5\% of world's population has disabling Hearing Loss (HL), defined as hearing loss 
greater 
than 40 [dB] in the better ear. One out of three adults aged over 65 years also are affected by disabling hearing loss. Current solution to 
address hearing loss is to compensate the approximate amount of loss in different frequencies, using a frequency-dependent amplification in a 
hearing aid \citep{Steinberg40,ZurekDelhorne87}. 
Yet hearing aid users complain about their ability for speech perception specially in environments such as restaurants where the background 
noise is similar to speech. Previous research supports the
hypothesis that HL, while a necessary factor, is not sufficient
in accounting for speech perception in Hearing Impaired (HI) ears
\citep{Plomp79b,Plomp86,YoonEtAl12,TrevinoAllen13b,AbavisaniAllen17}.

Various insertion gain prescription methods have evolved, such as National
Acoustics Lab (Revised) (NALR) \citep{Dillon01}, with the assumption that
the optimal insertion gain will improve audibility and as a result,
speech intelligibility. Although this gain treatment can help improve speech intelligibility for some HI ears, it has been shown that it can 
hurt 
speech intelligibility in nearly 12\% of cases \citep{AbavisaniAllen17}. The persistence of speech loss, once audibility has been
compensated, supports the possibility that there must be other factors,
such as outer hair cell loss, that are playing an important role in HI speech recognition. A speech metric that provides diagnostic
information would be easily justified, but to date, such speech metrics
have not been successful.

One major problem with focusing on audibility is that there has been
no fundamental understanding of the precise nature of the speech cues,
namely, which speech features need to be audible? The popular view of
speech cues are \emph{distinctive features} such as \emph{voicing, manner,
place and nasality} \citep{MillerNicely55}. These broad-brush features
are production rather than perception based, thus they do not account for
the large within-class variability, as they do not vary within a class, it is impossible for them to account for within-class variability. 
\citep{ToscanoAllen14}. Acoustic features that are necessary
for Normal Hearing (NH) listeners are also necessary for HI listeners,
but they may not be sufficient \citep{TrevinoAllen13a}. Consistent
token-specific confusion groups between HI listeners support the
hypothesis that HI ears use similar cues, despite the audiometric
configuration \citep{TrevinoAllen13b,AbavisaniAllen17}, but no theory exists that can clearly identify what these cues may be.

It was shown in a number of earlier studies that the errors HI ears
make depend on the token, not just on consonant or feature classes 
\citep{TrevinoAllen13a,TrevinoAllen13b}. These studies showed that our
traditional view of class-average errors can be misleading. 
At any amplification condition, there are numerous zero-error tokens along with a few high error tokens, and 
averaging, hides the degree of error for individual errorful tokens, thus diminishes the judgment of 
received benefit from that amplification procedure. To identify the errorful tokens, we need to look into error changes at various conditions 
and keep the token index fixed, as suggested by 
\citet{AbavisaniAllen17}. To do so, we may look into the accumulated error differences to evaluate improvement or degradations across 
various amplification gain treatments.

\begin{figure}[htbp!]
\centering
\includegraphics[width=\textwidth]{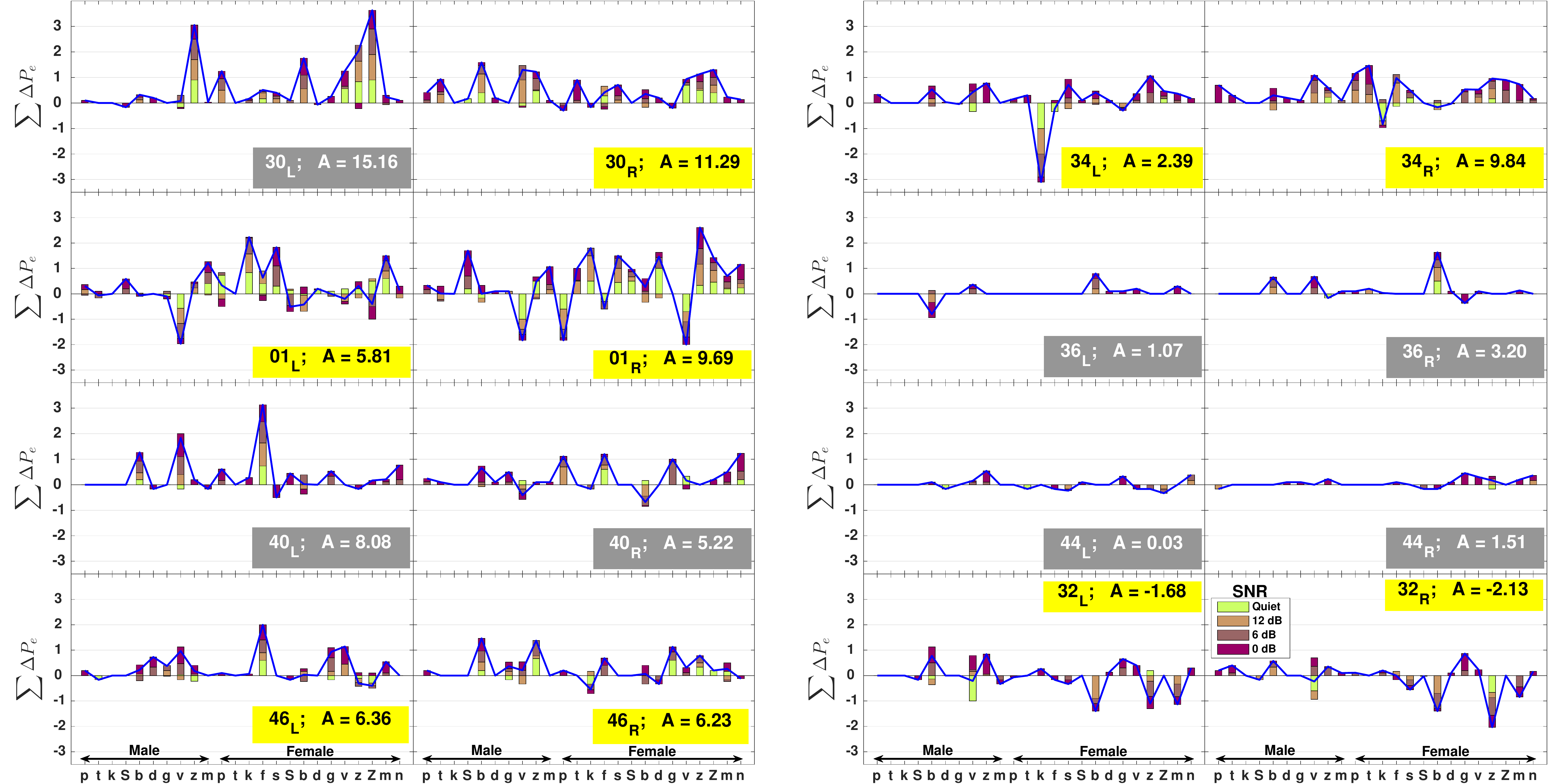}
\caption{\footnotesize Accumulated error differences ($\sum\Delta P_e$) for each subject; the line shows the difference between improved and 
degraded tokens error. Abscissa shows the 24 male and female talker/consonants (the vowel \textipa{/A/} is omitted to save space, and 
consonants 
\textipa{/S,Z/} are shown as /S,Z/). (A) in each panel shows the area under the $\sum\Delta P_e$ curve and it is an overall measure on 
helpfulness of frequency dependent treatment amplification versus a flat gain amplification \citep{AbavisaniAllen17}.}
\label{f:DeltaPe}
\end{figure}

Figure \ref{f:DeltaPe} shows an example on how one can evaluate a frequency dependent hearing aid amplification comparing to a 
flat gain amplification, and identify the tokens in which the treatment hurt the phone recognition by HI ear. This figure also shows that 
for a given HI ear at a given condition, out of many test tokens, there are only a few errorful tokens that need treatment to reduce 
recognition error.

This suggests the need to look deeper into individual differences,
to get a better understanding of how HI ears recognize speech. For a given HI ear, it is difficult to predict
which tokens can be correctly recognized, and which cannot, as they are different for each ear. To advance
understanding of this idiosyncratic deficiency of HI ears, a more
sensitive test is required. Such test may include pre-evaluated tokens with a perceptual measure, to control token dependent variability in 
speech perception for HI ear during the test.

In normal hearing ears each consonant becomes masked at a token dependent
threshold, denoted \SNRninty. The \SNRninty~is defined as the SNR in which on average, NH ears can recognize the token at least with 90\% 
correctly (the score is 90\% on average for NH ears).
As the noise is
increased from Quiet (no noise), the identification of
most sounds goes from less than 0.5\% error to 10\% error (at \SNRninty), and then
to chance performance, over an SNR range of
just a few [dB] (i.e., less than 10 [dB]) \citep{ToscanoAllen14}. Hence
\SNRninty~is an important token-specific threshold metric of noise
robustness, that may be used as the perceptual measure for the token.

\begin{figure}[htbp!]
\centering
\includegraphics[width=.3\textwidth]{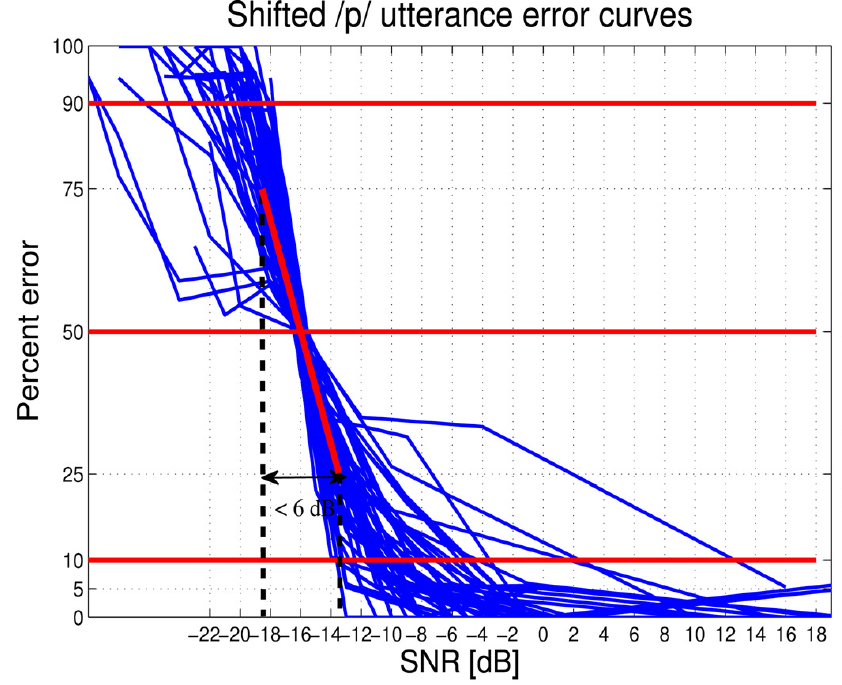}
\caption{\label{f:snr50} \footnotesize Individual \textipa{/p/} error curves aligned at their 50\% error values. The solid line shows the 
average master error curve, which falls from 75\% to 25\% error over 6 [dB] \citep{SinghAllen12}.}
\end{figure}

Previous studies showed that by examining many tokens of a particular Consonant-Vowel (CV) sound, in various noise conditions, in NH speech 
recognition experiments, one may construct the procedure to detect the \SNRninty~perceptual measure. Fig.~\ref{f:snr50} illustrates the 
results 
of such experiment for various tokens of \textipa{/p/} sound, where their error versus SNR curves are shifted to align them on the 50\% 
recognition point (SNR$_{50}$). According to Fig.~\ref{f:snr50}, if the amount of noise is increased, the score would drop significantly 
(over 50\%) for most of the sounds in just a few [dB] (i.e., 6 [dB]). This constitutes the theorem that for NH listeners, testing speech 
tokens at a few [dB] higher noise would drop the score significantly. Conversly, testing tokens at noise levels well above the 
\SNRninty, will result in correct recognition for NH listeners \citep{SinghAllen12}.

Accordingly, to test the idiosyncratic phone recognition for HI ears for a particular consonant, one should select the subset of tokens that 
have sharp score drop passing the \SNRninty~threshold. Fig.~\ref{f:snr50} shows that most of tested tokens fall within this criterion. By 
definition of \SNRninty~as a perceptual measure, NH listeners should have similar scores on tokens with similar \SNRninty~if tested at noise 
levels above the \SNRninty. Additionally, 
comparing two tokens with well separated \SNRninty~(i.e., $|\Delta$\SNRninty$|\geq$ 6 [dB]) at SNR equal to the higher \SNRninty~of two tokens
(i.e., at the \SNRninty~of less salient token), NH scores should vary significantly. By testing HI ears at noise levels much less than the 
threshold \SNRninty~of the token, we propose to quantify the idiosyncratic behavior of 
HI ear comparing to NH ears. The background noise level may be measured by methods such as the one explained in \citet{Lee07}.

Previous studies reveal that hearing loss can cause confusions for consonants where the primary cue region is within the hearing loss 
frequencies \citep{AbavisaniAllen17,ColePhD17}. This phenomenon is confirmed also on NH ears by high/low pass filtering of the primary cue 
region 
\citep{LiMenonAllen10,LiAllen11a}. Furthermore, it is confirmed that amplification of primary cue region improves recognition score for both 
NH and HI ears \citep{KapoorAllen12,ColePhD17}.

One research direction yet to be explored, is the question of whether varying the frequency of 
primary cue region can improve the speech recognition for HI ears. The very first step toward investigating on whether it is possible to 
improve the speech recognition score based on frequency component 
manipulations, will be to explore such effects in rather small scale such as what we call the ``\textit{frequency fine-tuning}" of the primary 
cue region. For fine-tuning, we may impose a change on the time-frequency window that is necessary to identify the consonant. One way 
to implement such a change is to vary the vowel in the token and keep the consonant and SNR the same. From previous studies conducted by 
\citet{LiAllen11a,SinghAllen12,Regnier08a}, we know that the intelligibility of a token in noise, is correlated to its \SNRninty~and also 
that the \SNRninty~is correlated with the relative intensity of the primary cue of the token. Thus, to keep the intelligibility of the primary 
cue in a similar level as the original token, we should select a new token that has a similar \SNRninty~to the original token.

Changing the vowel has a number of effects on NH consonant recognition, including changing the center frequency of 
the burst spectrum \citep{Winitz72}, changing the formant transitions \citep{Ohman66,Delattre55,Sussman91}, changing the acoustic 
spectrotemporal context within which the listener tries to 
identify the relevant cues \citep{Lisker75}, and changing the set of valid English words that are activated by the CV pair (an 
effect that has been shown to change the threshold for correct consonant recognition) \citep{Ganong80}. By keeping the NH-based perceptual 
measure (\SNRninty) the same across changing the vowel for HI listeners, and by testing at SNRs well above the \SNRninty~threshold, we control 
the approximate intensity of the primary cue region, which has the dominant effect on the intelligibility of token. Thus, if vowel change 
increases the error, candidates for such loss would be the HI audiometric configuration in conjunction with the vowel change effects on NH 
listeners.

To investigate the role of \SNRninty~perceptual measure in the improvement of HI consonant recognition, one may replace the errorful CV with a 
new but more intelligible CV with the same consonant and vowel. Being more intelligible is quantified in terms of the \SNRninty~of the token. 
Thus, it is proposed to replace the less salient token (higher \SNRninty) with a more salient token (lower \SNRninty) with a new talker, to 
increase the score for HI phone recognition. Given that the amount of noise in the experiment is much less than the \SNRninty~threshold for 
both tokens, such talker change should reduce error for HI ears, unless there are other factors involved. One candidate for such unexpected 
error path would be the conflicting cues becoming more available in the new token or noise condition.

In this article, we first explain the adaptive testing procedure to collect consonant recognition data from HI listeners. Then we provide 
preliminary results of experiment where the perceptual measure \SNRninty~varies for same CV sound, and the experiment where the perceptual 
measure \SNRninty~was kept in similar level, but the vowel changed for the same consonant. Finally, we discuss cases where such intervention 
went in the opposite direction as expected. These tests are directed at the fine-tuning of hearing aid insertion gain, with the ultimate goal 
of improving speech perception, and to precisely identify when and for what consonants HI ear needs treatment to enhance speech 
recognition.

\section{Methods}

Since we are interested to investigate the speech perception for HI ears in situations similar to real world experience, we need to design 
experiments that test HI speech recognition in speech-weighted noise. To explore the role of noise in such experiments, 
this study proposes to use the speech tokens at four SNR levels well above
\SNRninty\ (i.e., all SNRs should be above \SNRninty+6 [dB] for each token). With such a scheme, a single error
is highly statistically significant, since for the NH ear, one error in 
40 presentations at \SNRninty+6 [dB] is rare \citep{SinghAllen12}. Therefore, such schedule
is highly efficient in characterizing each HI ear, to determine what are the errorful tokens, and which consonants are problematic for HI 
ears to recognize. Previous studies show that HI listeners will have errors in recognizing tokens for only a subset of tokens 
(a few tokens out of all the presented tokens), if the tokens are presented well above their 
\SNRninty~\citep{AbavisaniAllen17,TrevinoAllen13b}. Once high error sounds
have been identified, one may seek the optimum treatment (insertion gain) to efficiently prevent increase of the token error relative 
to flat gain condition, for those errorful tokens.

\subsection{Speech materials}

Throughout these studies, the term token refers to one of the specific consonant-vowel (CV) sounds. We planned to test a wide range of different 
consonants to cover plosive, fricative and nasal sounds. The tested consonants are \textipa{/p, t, k, f, s, S, b, d, g, v, z, Z, m, n/}. These 
consonants combined with vowels \textipa{/A, \ae, I, E/} form the CV speech 
database for this experiment. For each CV sound there are two instances assembled in two sets: set T$_1$ which includes CV sounds with 
\SNRninty~perceptual measure below -2 [dB], and set T$_2$ which includes same CV sounds with different talkers that are more salient and have 
\SNRninty much less than corresponding CV in set T$_1$ (i.e, $|\Delta$\SNRninty$|\geq$6 [dB] for same CV from sets T$_1$ and T$_2$).

The CV tokens were drawn from an earlier experiment that measured
the confusions as a function of SNR for 30 NH listeners
\citep{LiMenonAllen10}.  The tokens were restricted to be noise-robust,
defined as having a recognition error as measured by 30 NH ears of less
than 10\% at SNR = -2 [dB], with an average error of $<$3.1\% (i.e., less
than 1 in 32 trials \citep{PhatakAllen07a,SinghAllen12,ToscanoAllen14} at
the four test SNRs (i.e., 0, 6, 12 [dB] and Quiet).  During the testing,
speech shaped computer generated Gaussian noise was added to the token
at one of the four SNRs.

Each token was naturally spoken as an isolated (i.e., no carrier phrase)
consonant-vowel (CV) token, by an American English speaking talker, from a pool of eight female talkers and twelve male talkers, 
available from the Linguistic Data Consortium Database (LDC-2005S22)
\citep{Fousek04}.  The sampling rate was 16 [kHz]. 

The speech was presented at each subject's most comfortable level (MCL),
as determined during initial trials used to familiarize the subjects
with the task. Initially, the software was calibrated to present speech stimuli at 75 [dB SPL]. 
The subjects were allowed to subsequently
adjust the presentation level at any time during the experiments, and if they did such adjustment, the new presentation level is saved in the 
experiment log file.

\subsection{Subjects}
\label{s:exp1sub}
The target subjects for these experiments are native English speakers who have mild to moderate hearing loss with the age between 18-64 years. 
These subjects are recruited from Urbana-Champaign, IL, community. IRB approval was obtained from the University Review Board. Subjects were 
paid. All subjects had hearing loss greater than 20 [dB] for at least one frequency in the range 0.25-4 [kHz].

\begin{figure}[htbp!]
\centering
\includegraphics[width=.3\textwidth]{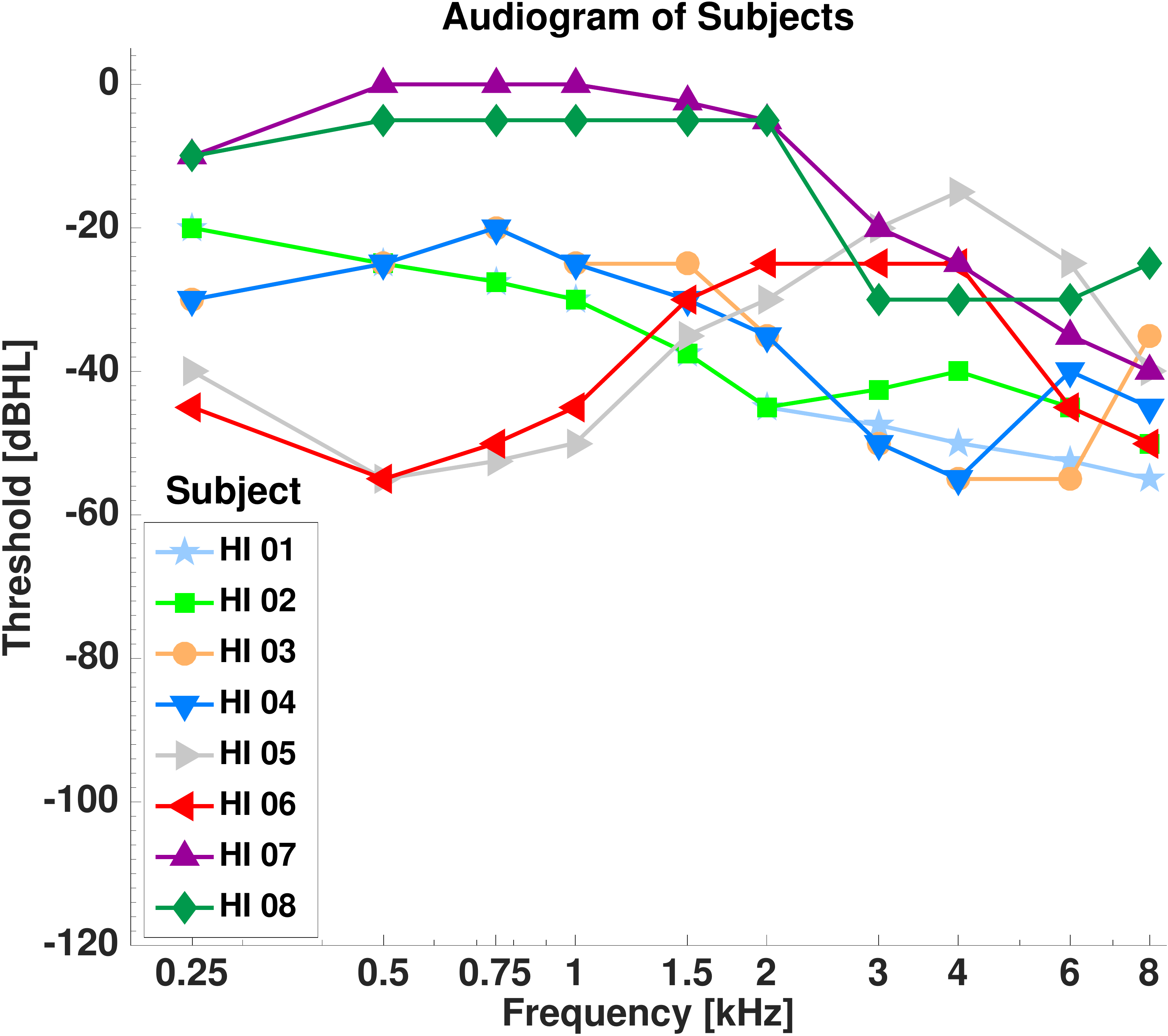}
\caption{\label{f:PTT} \footnotesize Pure Tone Thresholds (PTT) of the subjects participated in current study. Consecutive subject numbers 
corresponds to the 
same person.}
\end{figure}

Figure \ref{f:PTT} illustrates the pure tone thresholds of the subjects whose test results appear in current study. All subjects have 
mild to moderate hearing loss in high frequencies. In addition, subjects HI$_1$, HI$_2$, HI$_3$ and HI$_4$ have mild hearing loss in low 
frequencies. Subjects HI$_5$ and HI$_6$ (same person) have moderate hearing loss in low frequencies as well. All the pure tone thresholds have 
been evaluated within the past year prior to the experiments.

\subsection{Experiment design}

To investigate the effect of cue enhancement, all the conditions were the same, other than 
the talker (with same gender) which is replaced by a talker who produced the target consonant more clearly 
in terms of \SNRninty. The CV remains the same. Additionally, to investigate the effect of frequency fine-tuning of consonants via changing the 
vowel, all the conditions were the same, other than the talker (with same gender) and vowel of token which is replaced by a token with similar 
salience in terms of \SNRninty. The consonant remains the same.

\begin{figure}[htbp!]
\centering
\includegraphics[width=\textwidth]{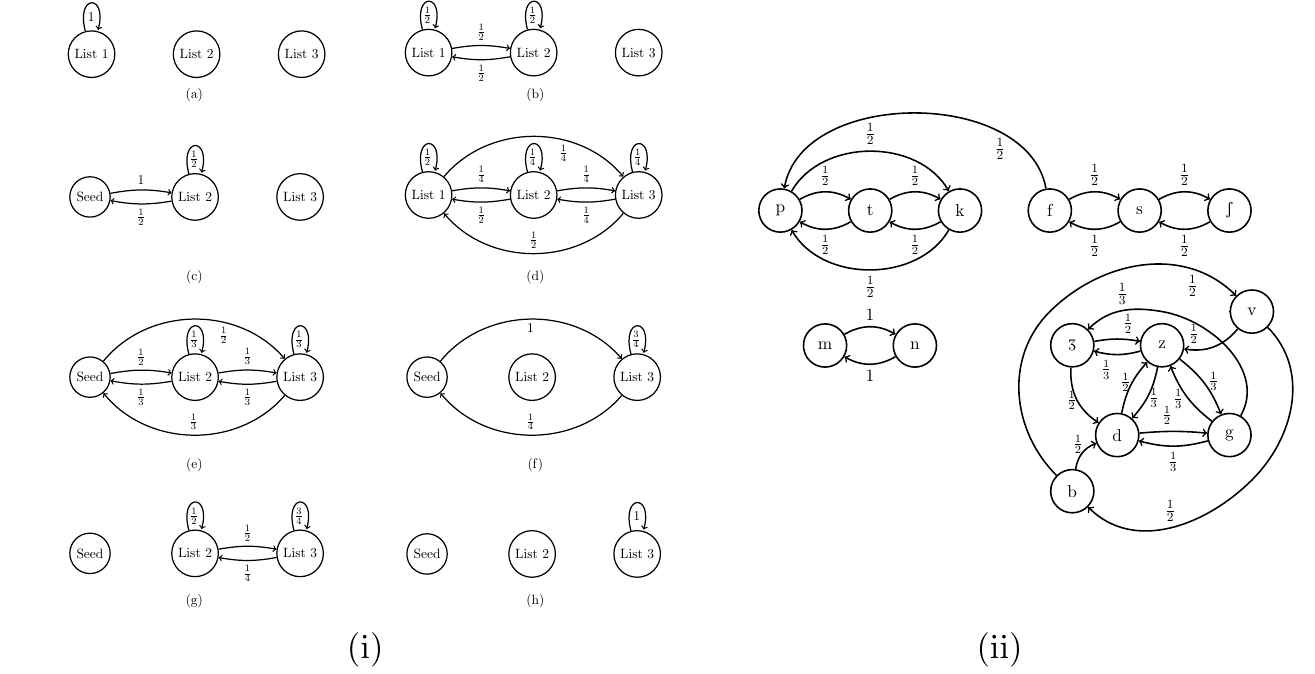}
\caption{\label{f:Lists} \footnotesize Scheduled procedure for adaptive testing. Numbers on each edge show the probability of transition: (i) 
Transition probabilities between the lists at different stages of the experiments: (a) Initial probability distribution 
indicating that only List 1 contains tokens, (b) distribution when only List 1 and 2 contain tokens to present, (c) distribution 
when only List 2 contains tokens, (d) distribution when all lists 
contain tokens, (e) distribution when only List 2 and 3 contain tokens, (f) distribution when only List 3 contains tokens, (g) distribution when 
only List 2 and 3 contain tokens and List 3 includes more than than 10 distinctive consonants, and (h) distribution when only List 3 contains 
tokens and it has more than 10 distinctive consonants. (ii)
Transition probabilities between various consonants, which are used to add induced confusing consonants as seed tokens during the experiments.}
\end{figure}

The experiment starts on List 1 (see Fig.~\ref{f:Lists}(a)) with tokens 
including both male and female talkers for the 14 available consonants associated with vowel \textipa{/A/} at SNR = 0 [dB]. These starting 
tokens are already highly intelligible as they all have \SNRninty~below 0 [dB]. Thus NH listeners should recognize them correctly. If the HI 
ear has error for a token, that token will be presented two more times in List 2: once at 0 [dB] and once at 6 [dB] (one level higher SNR). 
After these three presentations, if the HI ear has two errors out of three, we consider the token to be a susceptible token that needs more 
scrutiny. Hence, such token will be moved to List 3, where it will be presented 10 times at each SNRs of 0, 6, 12 [dB] and 
Quiet, making a total of 40 more presentations.

As soon as a token reaches List 3, other versions of that CV will be added to List 2, in order to investigate the enhanced cue 
effects (more salient talker). Also, other versions of the same consonant with various vowels will be added to List 2, such that for each 
CV, there will be at least two different talkers one with similar \SNRninty~as the original CV, and one with better \SNRninty~(more salient). 
Furthermore, to prevent subjects from guessing the correct response, seed tokens that are confusable with the original consonant will be added 
to List 2. The confusable consonants are determined from previous CV recognition experiments in noise with NH listeners \citep{MillerNicely55}. 
Fig.~\ref{f:Lists}(b) illustrates the confusable consonants and their transition probabilities.

Token orders are randomized in all experiment lists initially and whenever a new token is added to a list. Since List 2 and 3 usually include 
more tokens, seed tokens are provided in a Seed List to mix the token presentation order with seed tokens and increase the randomness.
Additionally, presentation from different lists are randomized to prevent subjects from guessing. The transition probabilities between lists are 
shown in Fig.~\ref{f:Lists}(a).

A Matlab$^\circledR$ graphical user interface was provided to run the experiment. All of the data collection sessions were conducted with the 
subject seated in a single-walled, soundproof booth with the door of the outer room closed. The speech was presented through an Etymotic ER-3 
insert ear phone, one ear at a time. The contra-lateral ear was not masked or occluded. To familiarize the subjects with the testing paradigm, 
a practice session was run using non-test tokens. The MCL was 
determined during the practice session. 

After hearing each token, the subject was instructed to choose the response from 14 possible consonant labeled buttons that were provided on 
the screen via a graphical interface. To get more precise results, subjects were allowed to play uncertain tokens up to two additional times 
before making their decision. To reduce fatigue, subjects were encouraged to take short breaks approximately every 20 min.

\subsection{Data Analysis}

Collected data were saved into log plain text files on disk. For each presented token, the saved data include talker, played consonant, 
vowel, heard consonant, SNR, Sound Pressure Level (SPL), number of repeats, List number, name of wave sound file, and the time the subject took 
from hearing the CV till hitting the response button. 

From the collected data, we use the responses from List 3, which are the results of full investigation of susceptible tokens, that are 
presented evenly at four SNRs (0, 6, 12 [dB], and Quiet). From these data, we can form the \textit{confusion matrix} as a function of SNR. Since 
we conduct the study on 14 consonants, the confusion matrix will be of size 14$\times$14. Each of the tokens presented in List 3, has an 
empirical probability distribution defined by a row of the count (unnormalized confusion) matrix. We refer to the $i^{th}$ token as CV$_i$, 
$i=1,\dots14$. The probability of error of this token is:
\begin{equation}
\label{eq:Pe}
P_e(CV_i,SNR) = \sum_{j\neq i}P\{heard CV_j | spoken CV_i\},
\end{equation}
where $P_{ii} = 1 - P_e$ is the corresponding probability of correct response (diagonal element). For simplicity in notation, we may refer to 
$P_e(CV_i,SNR)$ as $P_e$. Given the above probability of error for each of the tokens, the average error of erroneous consonants for each ear is 
then
\begin{equation}
\label{eq:avgPe}
\overline{P_e}(Ear,SNR) = \frac{1}{N_3}\sum_{i=1}^{N_3} P_e(CV_i,SNR),
\end{equation}
where $N_3$ is the number of CV tokens that are reached to List 3, i.e., the number of consonants that were hard to hear.

Another measures that is considered is the confusion pattern (CP); for a given token, the confusion pattern is a plot of one row of the 
confusion matrix (i.e., $P_{heard|spoken}(SNR)$), as a function of SNR \citep{Allen05b}. This measure shows how the token score and confusions 
depend on SNR.

\section{Results}

In this section we discuss the results of both experiments on cue enhancement and frequency fine-tuning of the speech cue for HI listeners. 
While on average, HI ears responded positively to cue enhancement, the result of vowel change is mixed and varies regarding various vowel. 

\subsection{Cue enhancement}

\subsubsection{Error summary}

\begin{figure}[htbp!]
\centering
\includegraphics[width=.7\textwidth]{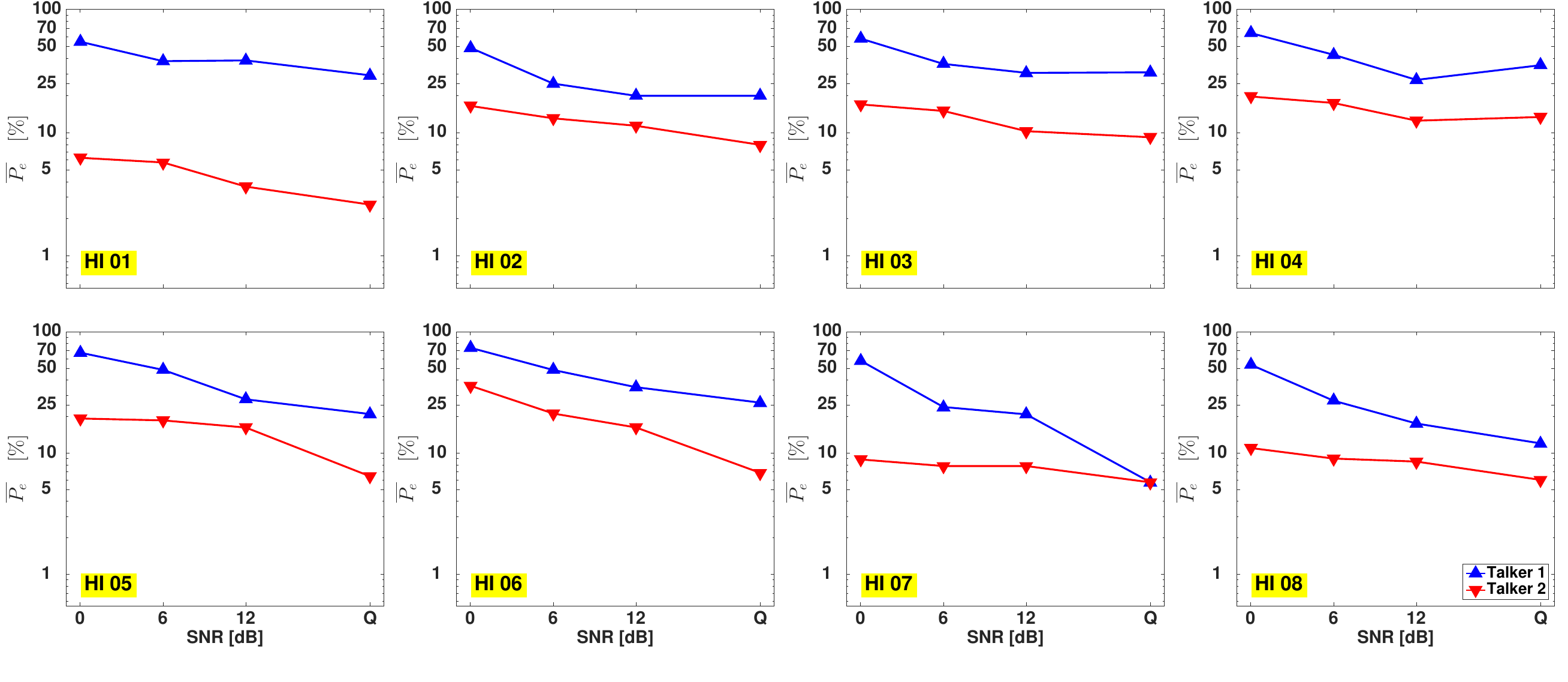}
\caption{\label{f:AvgT} \footnotesize Average probability of error in log scale versus SNR; in each 
panel, the blue curve shows the average error regarding the talkers with higher \SNRninty~(less salient), and the red curve corresponds to the 
talkers with lower SNR90 (more salient).}
\end{figure}

Figure \ref{f:AvgT} illustrates the average log-probability of error $\overline{P_e}(Ear,SNR)$ for all HI ears in current study. 
From Fig.\ref{f:AvgT} we observe that $\overline{P_e}$ have linear relationship with SNR for HI ears. Generally, when the nosie decreases, the 
error also decreases. However, there are cases such as subject HI$_4$ (top right panel in Fig.~\ref{f:AvgT}), where eliminating noise from 
SNR = 12 [dB] to Quiet, caused the error to increase. This happened for tokens from both set T$_1$ talkers and the more salient talker set 
T$_2$, for this subject. Such phenomenon indicates that conflicting cues became available when the nosie reduced, causing the subject to 
confuse the consonant with the corresponding consonant from the newly available cues.

As depicted, on 
average, when tokens are replaced by same CV but with better perceptual measure (lower \SNRninty), all HI ears responded with better consonant 
recognition. This confirms that HI ears use same perceptual features as NH listeners and if one enhances the speech cue, the speech perception 
will improve for HI listeners.

Given that in both experiments no frequency-dependent amplification was provided for HI ears, one may expect the two $\overline{P_e}(Ear,SNR)$ 
curves before and after replacing token with the same CV but with more salient talker, should be parallel. Fig.~\ref{f:AvgT} confirms such 
expectation for most HI ears, however, passing from SNR = 12 [dB] to Quiet, there are cases where these two curve converge (subject HI$_7$) or 
diverge (subject HI$_6$). Accordingly, at the presence of noise $\overline{P_e}(Ear,SNR)$ curves converged for subjects HI$_2$, HI$_5$, HI$_7$ 
and HI$_8$, when the talker is replaced to enhance primary cue. This convergence is an indication of the limit on cue enhancement (on average) 
for consonant recognition improvement for HI ears. Apparently, some HI ears will have errors for some tokens even though the cue is enhanced 
and the nosie is reduced. We should look into individual consonant error changes to identify corresponding consonants that are not responding 
to cue enhancement for each HI ear. Such consonants are idiosyncratic for each HI ear.

\subsubsection{Improvement and degradation due to talker change}

\begin{figure}[htbp!]
\centering
\includegraphics[width=.7\textwidth]{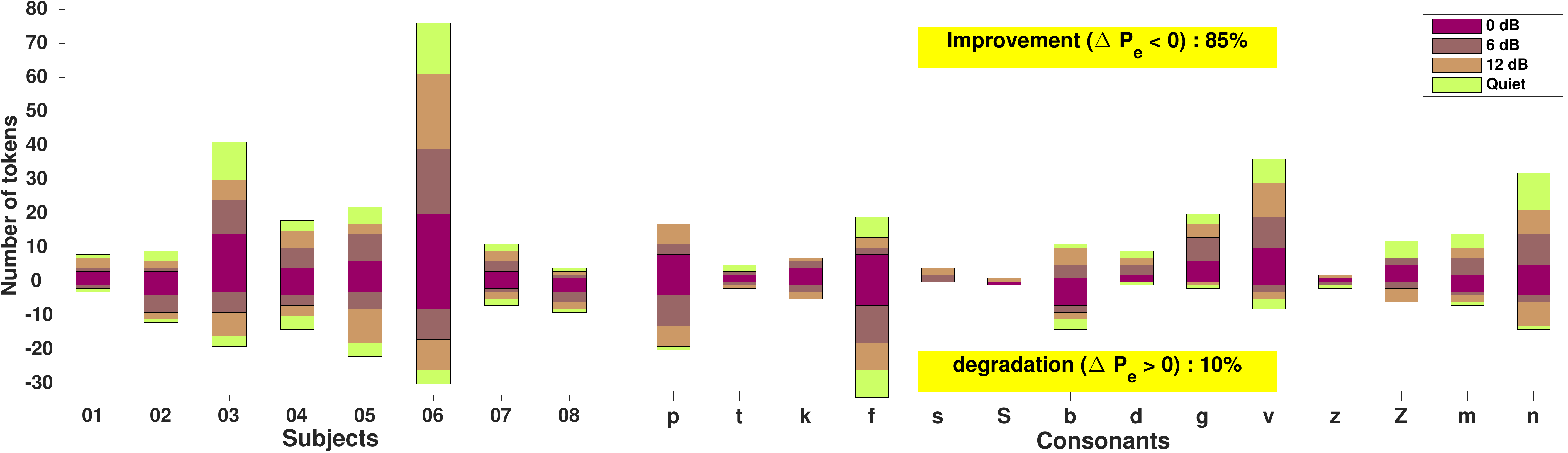}
\caption{\label{f:Timpdeg} \footnotesize .}
\end{figure}

To have more precise understanding on the effects of cue enhancement through changing the talker to a more salient talker, we may look into 
the individual token error changes. Fig.~\ref{f:Timpdeg} illustrates the token error changes for subjects (left panel) and for 
consonants (right panel). The ordinate is the number of tokens that are improved or degraded due to replacing talker with another talker that 
had better \SNRninty. These improvement and degradations are from the consonants which had error both when provided token from set 
T$_1$ (less salient talker) and set T$_2$ (more salient talker), at various SNR. Overall, considering the tokens where improving 
\SNRninty~vanished the error, 85\% of tokens are improved and 10\% of tokens are degraded.

According to the degradations by subjects, we observe that subjects HI$_6$, HI$_5$, HI$_3$ and HI$_4$ had the most 
degradations, respectively. By comparing the audiometric thresholds (Fig.~\ref{f:PTT}), we find out that these subjects had hearing loss in 
low frequencies (below 1 [kHz]) in addition to high frequency sloping loss. 

Moreover, right panel in Fig.~\ref{f:Timpdeg} shows that consonants \textipa{/f, p, n/} had the most degradations, respectively. Analysis such 
as the one in Fig.~\ref{f:DeltaPe} will describe consonant degradations according to HI ears, thus one may associate the specific consonant 
degradation with audiometric configuration.

\subsection{Frequency fine-tuning}

\subsubsection{Error summary}

\begin{figure}[htbp!]
\centering
\includegraphics[width=.7\textwidth]{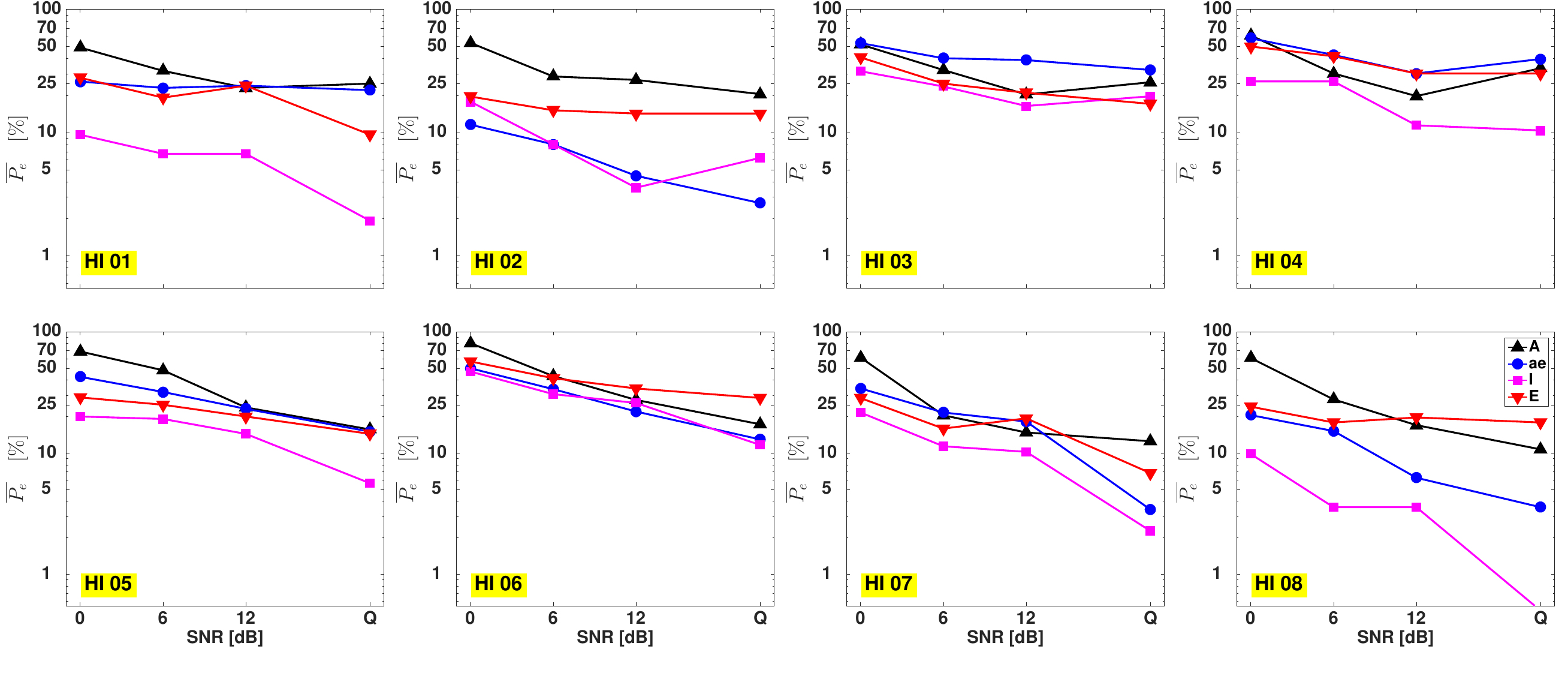}
\caption{\label{f:AvgV} \footnotesize Average probability of error for tokens in vowel change experiment; in each panel vowels 
\textipa{/A, \ae, I, E/} are shown by symbols $\blacktriangle$, {\large$\bullet$}, $\blacksquare$~, {\small$\blacktriangledown$}~, 
respectively. 
In the legend, vowels \textipa{/A, \ae, I, E/} are shown with characters \textit{/A, ae, I, E/}, respectively.}
\end{figure}

Figure \ref{f:AvgV} illustrates the $\overline{P_e}(Ear,SNR)$ for each subject when the consonant is kept the same and the vowel is 
changed while the perceptual measure \SNRninty~of both tokens are kept in the similar level. As expected, the $\overline{P_e}$ has a linear 
relationship with SNR in all cases with various vowels. Following, we describe several events that are observable from these average error vs 
SNR plots. 

First, it appears that some subjects had less average error for vowel \textipa{/I/} comparing to other tested vowels. By checking the average 
\SNRninty~of the initial test tokens in table \ref{t:Vimpdeg}, we observe that on average, tokens with vowel \textipa{/I/} had lower 
\SNRninty. Second, subject HI$_4$ (top right panel) had higher error in Quiet than at SNR = 12 [dB] for various vowels with the exception of 
vowel 
\textipa{/I/}. Third, in the presence of noise, the $\overline{P_e}$ curve remains parallel for various vowels for most of the subjects, 
meaning 
that reducing the noise improved CV recognition scores with similar degrees. There are exceptions to this observation such as subjects HI$_7$ 
and HI$_8$ (same person) for vowel \textipa{/E/} (marker {\small$\blacktriangledown$} in lower right two panels in Fig.~\ref{f:AvgV}). 

Fourth, for several vowels $\overline{P_e}$ curve does not decay as the noise decreases. These include vowel \textipa{/\ae/} for subjects 
HI$_1$ and HI$_3$ (marker {\large$\bullet$}) and vowel \textipa{/E/} for subjects HI$_1$, HI$_2$, HI$_5$, HI$_7$ and HI$_8$
(marker {\small$\blacktriangledown$}). Fifth, all subjects are not sensitive to eliminating the noise for at least one vowel, meaning that 
the $\overline{P_e}$ curve did not change across SNR = 12 [dB] and Quiet. For some subjects, there are vowels in which the $\overline{P_e}$ 
did 
not change significantly from SNR = 0 [dB] (highest noise) to Quiet. These cases include vowel \textipa{/\ae/} for subjects HI$_1$ and HI$_3$, 
and vowel \textipa{/E/} for subjects HI$_2$ and HI$_8$.

All of these observations need to be investigated by more detail combining subjects' audiometric configuration and individual consonants 
contributing to such events, so we can have better judgement on whether the coarticulatory cues that affect NH phone recognition, also affect
HI phone recognition by similar weights.

\subsubsection{Improvement and degradation due to vowel change}

\begin{figure}[htbp!]
\centering
\includegraphics[width=\textwidth]{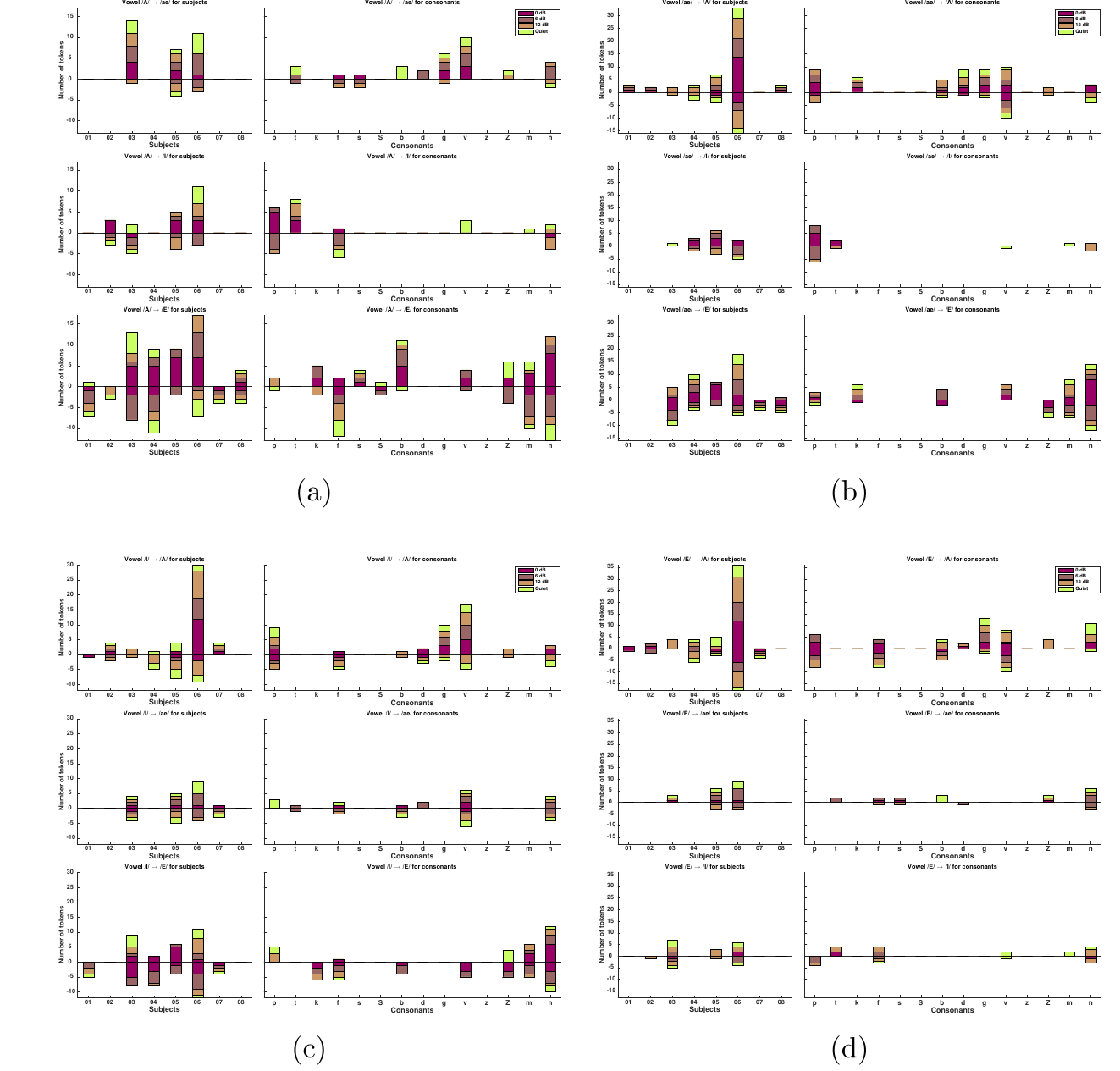}
\caption{\label{f:impdeg} \footnotesize Number of improved versus degraded tokens when the vowel was replaced, and a new token with similar 
\SNRninty~is presented to HI listeners: (a) Vowel \textipa{/A/} replaced by 
\textipa{/\ae, I, E/}, (b) Vowel \textipa{/\ae/} replaced by \textipa{/A, I, E/}, (c) Vowel \textipa{/I/} replaced by \textipa{/A, \ae, E/}, 
and (d) Vowel \textipa{/E/} replaced by \textipa{/A, \ae, I/}. On the title in each panel, vowels \textipa{/A, \ae, I, E/} are shown with 
characters \textit{/A, ae, I, E/}, respectively. Consonants \textipa{/S, Z/} are shown by \textit{S, Z} in the abscissa labels. The cases where 
changing the vowel vanished the error are not shown in these plots.}
\end{figure}

Table \ref{t:Vimpdeg} shows the percent improvement and degradations when the vowel is changed from the original vowel, and the token is 
replaced by a new token with same consonant and different vowel, but with similar \SNRninty~that reflects similar perception for an NH 
listener. The average \SNRninty~of the tokens for each vowel is also provided in table \ref{t:Vimpdeg}.

The percentage in table \ref{t:Vimpdeg} indicate the absolute number of improved/degraded tokens and not the degree of 
improvement/degradation. To find out about the degree, methods such as the one used in Fig.~\ref{f:DeltaPe} is needed. Overall, vowels 
\textipa{/A, \ae, E/} improved and degraded similarly. Their average \SNRninty~perceptual measure is also very close.

Figure \ref{f:impdeg} illustrates the overall improvement vs degradations when the consonant is kept the same and the vowel changed to a new 
vowel, excluding the cases where the error vanished by vowel change. Each panel include two summary bar plots: the left axes indicates the 
improvement vs degradation for subjects, and the right axes indicates same information collapsed on various consonants. 

At a glance, Fig.~\ref{f:impdeg} shows that subject HI$_6$ had the most number of improvement specifically when vowels \textipa{/A,E/} are 
involved. This can be related to this subject's low frequency hearing loss that affects low frequency energy vowels such as \textipa{/A,E/}. 
On the other hand, looking to the degradations, we see that subject HI$_6$ had the highest degradations when the vowel changed from 
\textipa{/I/} to other vowels. Generally, vowel \textipa{/I/} has low frequency first formant along with a high frequency second formant 
\citep{Patterson82,Hillenbrand95}. Thus, the low frequency hearing loss of subject HI$_8$ can play a role in degradation of perceiving  
consonants+\textipa{/I/}, comparing to other CVs. Such conclusions may be deemed for each individual subject by accompanying their hearing 
loss with the frequency components of the degraded vowels.

\begin{table}[htbp!]
	\centering
	\caption{\label{t:Vimpdeg} Percent improvement and degradations in errors on HI consonant recognition when the vowel changed.}
\begin{tabular}{cccc}
	\textbf{Changed Vowel} & \textbf{Improvement [\%]} & \textbf{Degradation [\%]} & Average \SNRninty\\ \hline
	\textipa{/A/} & 75 & 14 & -9.6 \\
	\textipa{/\ae/} & 71 & 16 & -9.9 \\
	\textipa{/I/} & 63 & 24 & -12 \\
	\textipa{/E/} & 72 & 18 & -10.5 \\
\end{tabular}
\end{table}

\section{Discussion}

\begin{figure}[htbp!]
\centering
\includegraphics[width=.5\textwidth]{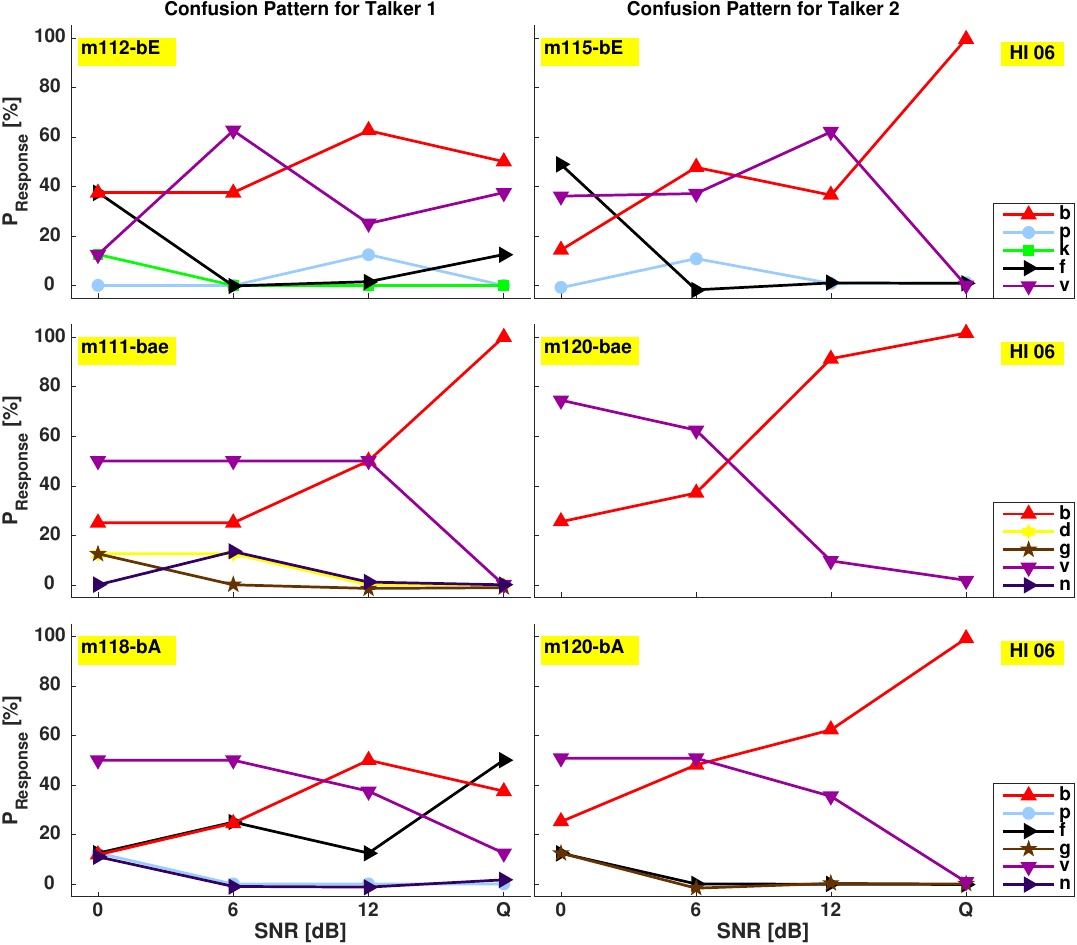}
\caption{\label{f:CP} \footnotesize Examples of confusion patterns from subject HI$_6$ for consonant \textipa{/b/} associated with 
vowels \textipa{/E, \ae, A/}, respectively, where some of the responses 
were unexpected; each panel shows the probability of response versus SNR for a different token that includes the same consonant. In each row, 
left panel shows confusion pattern for the CV with higher \SNRninty~(less salient) and right panel shows the confusion pattern for the same CV 
that is enhanced in terms of perceptual measure \SNRninty~(more salient). Panels from top to bottom show the change in confusion pattern 
as the vowel changed. Right panels in each row also include the legend that shows target consonant on top and all other responded 
consonants afterward. In the textbox in each panel, vowels \textipa{/A, \ae, E/} are shown with characters \textit{/A, ae, E/}, 
respectively.}
\end{figure}

It is clear from Fig.~\ref{f:AvgT} that on average enhancing the token in terms of perceptual measure (\SNRninty) will improve the speech 
perception for HI subjects. However, when looking into individual tokens, this may not be the case always. Fig.~\ref{f:CP} illustrates the 
confusion patterns for several cases where the intervention (change in vowel or talker), showed some unexpected results for the HI listener. 
For the vowel change
experiments, the changes in error pattern depends to each subject's audiometric configuration so we cannot extend a rule from average error. 

\begin{figure}[htbp!]
\centering
\includegraphics[width=\textwidth]{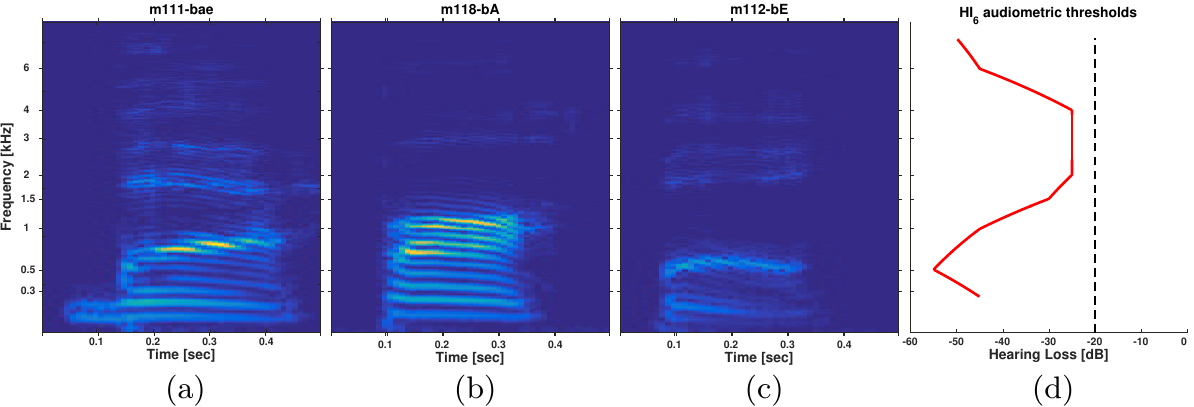}
\caption{\label{f:bV-Mel} \footnotesize Mel-frequency spectrogram of different tokens of consonant \textipa{/b/} at Quiet
along with the 
audiometric thresholds of subject HI$_6$ in Mel-frequency scale. Spectrograms illustrate token features on the same 
colormap scale. The dashed-line at -20 [dB] in the right panel (d) indicates the threshold of NH: (a) token \textipa{/b\ae/}, 
(b) token \textipa{/bA/}, (c) token \textipa{/bE/}, (d) the audiometric thresholds of subject HI$_6$.}
\end{figure}

The confusion patterns in Fig.~\ref{f:CP} informs about the role of cue enhancement (enhance \SNRninty) and the frequency fine-tuning in 
complex cases, where the speech perception for a consonant, did not follow the average rule. Figure \ref{f:CP} indicates the confusion pattern 
of perception of consonant \textipa{/b/} for subject HI$_6$. As it is shown, various 
\textipa{/b/}+vowel tokens had better score when the token is replaced with a better token in terms of \SNRninty. Comparing the score for 
less salient tokens with
different vowels (left panels in Fig.~\ref{f:CP}), for instance, we observe that at Quiet \textipa{/b\ae/} was recognized correctly, but 
\textipa{/bA/} and \textipa{/bE/} were perceived by 60\% and 50\% error, respectively. For explanation, we look into the spectrograms of 
these tokens provided in Fig.~\ref{f:bV-Mel}, along with the audiometric thresholds of subject HI$_6$.

\comment{
For example, Fig.~\ref{f:CP}(b) middle row shows 
that subject HI$_6$ with low frequency 
hearing loss, responded the to \SNRninty~enhancement of token \textipa{/gE/} with more confusion toward \textipa{/dE/} at SNR=12 [dB] and 
Quiet. But at SNR=6 [dB], this cue enhancement vanished the confusion with \textipa{/dE/} and the score went to 100\%. Similar phenomenon 
happened for subject HI$_4$ in perception of token \textipa{/vE/}, where by cue enhancement, the confusion with \textipa{/tE/} increased at 
SNR=12 [dB] and Quiet (see Fig.~\ref{f:CP}(c), top row). 
}

Since the characteristics of the labial stop consonant \textipa{/b/} is to have a diffuse spread of energy over a wide range of frequencies 
\citep{Blumstein79}, the strength of the burst release into the following vowel seems to have an important role in correct perception of 
this stop consonant for subject HI$_6$. Fig.\ref{f:bV-Mel}(a)-(c) show that the burst release of \textipa{/b/} at frequencies above 1 [kHz] 
is stronger in token 
\textipa{/b\ae/} comparing to \textipa{/bA/} and \textipa{/bE/}. This is evident by comparing higher frequency formants of these vowels in 
Fig.\ref{f:bV-Mel}(a)-(c). On the other hand, as depicted in Fig.\ref{f:bV-Mel}(d), subject HI$_6$ had better hearing abilities in 
mid-frequencies of 1.5-4 [kHz]. Hence, if the burst release of \textipa{/b/} associated with formants of the next vowel appears to be 
present in this frequency range, one would predict the correct perception. Fig.\ref{f:bV-Mel}(a) indicates that the second formant of vowel 
\textipa{/\ae/} falls in this range and is strong enough to be heard by HI$_6$. As appears in Fig.\ref{f:bV-Mel}(a)-(c), although the first 
and second formants 
of \textipa{/A/} and \textipa{/E/} are strong, but they fall into the sever HL range of subject HI$_6$. The first formant of \textipa{/\ae/} 
also falls in this range. Hence, we may conclude that the vowel formants in association with the audiometric configuration of the 
HI listener, have eminent role in consonant perception for HI listeners.

On the other hand, comparing the Voice Onset Time (VOT) of the vowel in the three panels in Fig.~\ref{f:bV-Mel}, we observe that the VOT for 
\textipa{/b\ae/} starts approximately 100 [msec] sooner than \textipa{/bA/} and \textipa{/bE/}. Thus, the VOT may also played a role in 
better perception of \textipa{/b\ae/} for subject HI$_6$. This indicates that the role of the VOT in the case of HI consonant recognition, 
is similar to the VOT role in NH consonant recognition as explained by \citet{Lisker75}.

The confusion patterns of Fig.\ref{f:CP} provides summary of some unexpected events as a result of frequency fine-tuning and cue enhancement 
for token. Each condition in these plots can be analyzed by illustrations such as Fig.~\ref{f:bV-Mel} for various tokens and subjects. This 
analysis indicated that for instance, given the audiometric thresholds of subject HI$_6$, for perception of \textipa{/b/}, a frequency 
fine-tuning toward vowel \textipa{/\ae/} would increase the score. Similar explanation may be given for other cases  
to recommend an individual frequency fine-tuning prescription for HI subjects.

\section{Conclusion}

Throughout this study, we analyzed the speech recognition data from HI listeners with mild to moderate hearing loss, to investigate the 
role of cue enhancement and frequency fine-tuning. The control factor in these experiments was the perceptual measure \SNRninty~which 
assures that in the SNR levels of the test study, NH listeners would recognize the tokens correctly. The results show that the cue enhancement 
with no frequency-dependent amplification improves consonant recognition for all HI subjects on average. 

The results of the frequency fine-tuning experiment, did not indicate any favorable vowel for consonant recognition on average. The CV tokens 
with vowel \textipa{/I/} that had slightly better \SNRninty, showed less improvement and more degradation when the vowel replaced, indicating 
the importance of the \SNRninty.

Using the confusion pattern plots such as the ones in Fig.~\ref{f:CP}, we can observe whether the enhancement in conditions such as noise, 
cue, frequency fine-tuning, does not provide the expected outcome. To further analyze the reason behind such event, we may look into the 
illustrations such as Fig.~\ref{f:bV-Mel}. The studied case showed that the strength of burst release of stop consonant \textipa{/b/} into 
the formants of the following vowel, plays an important role in identification of the consonant.

\section*{Acknowledgments}


\end{document}